\documentclass[prl,aps,twocolumn,showpacs,floatfix,superscriptaddress]{revtex4-1}
\usepackage{graphicx}
\usepackage{subfigure}
\usepackage{bm}
\usepackage{amsmath}
\usepackage{amssymb}
\usepackage[normalem]{ulem}
\usepackage[colorlinks,citecolor=black,urlcolor=blue,bookmarks=false,hypertexnames=true]{hyperref}
\usepackage{tabularx}
\usepackage{wasysym}
\usepackage{color}
\usepackage{array}
\usepackage{multirow}
\usepackage{makecell}
\usepackage[version=3]{mhchem}
\usepackage{float}
\usepackage{color}
\usepackage{lipsum}
\usepackage{ulem}
\usepackage{amsfonts}
\usepackage{fancyhdr}
\makeatletter
\newcommand{\colorcaption}[2][]{%
  \begingroup%
  \renewcommand{\@caption@fignum@sep}{ (Color online)  }%
  \caption[#1]{#2}%
  \endgroup%
  }
\makeatother
\makeatletter
\newcommand{\colorcaptions}[2][]{%
  \begingroup%
  \renewcommand{\@caption@fignum@sep}{ (Graphical abstract). }%
  \caption[#1]{#2}%
  \endgroup%
  }
\makeatother
\usepackage{xcolor}

\begin{document}
\title{Realistic flat-band model based on degenerate $p$-orbitals \\ in two-dimensional ionic materials}

\author{Jiang Zeng}
\thanks{Corresponding author: zengjiang@pku.edu.cn}
\affiliation{International Center for Quantum Materials, School of Physics, Peking University, Beijing 100871, China}

\author{Ming Lu}
\affiliation{Beijing Academy of Quantum Information Sciences, Beijing 100871, China}
\affiliation{International Center for Quantum Materials, School of Physics, Peking University, Beijing 100871, China}

\author{Haiwen Liu}
\affiliation{Center for Advanced Quantum Studies, Department of Physics, Beijing Normal University, Beijing 100875, China}

\author{Hua Jiang}
\affiliation{School of Physical Science and Technology, Soochow University, Suzhou 215006, China}

\author{X. C. Xie}
\affiliation{International Center for Quantum Materials, School of Physics, Peking University, Beijing 100871, China}
\affiliation{Beijing Academy of Quantum Information Sciences, Beijing 100193, China}
\affiliation{CAS Center for Excellence in Topological Quantum Computation, University of Chinese Academy of Sciences, Beijing 100871, China}


\begin{abstract}
Though several theoretical models have been proposed to design electronic flat-bands, the definite experimental realization in two-dimensional atomic crystal is still lacking. Here we propose a novel and realistic flat-band model based on threefold degenerate $p$-orbitals in two-dimensional ionic materials. Our theoretical analysis and first-principles calculations show that the proposed flat-band can be realized in 1T layered materials of alkali-metal chalogenides and metal-carbon group compounds. Some of the former are theoretically predicted to be stable as layered materials (e.g., K$_2$S), and some of the latter have been experimentally fabricated in previous works (e.g., Gd$_2$CCl$_2$).  More interestingly, the flat-band is partially filled in the heterostructure of a K$_2$S monolayer and graphene layers. The spin polarized nearly flat-band can be realized in the ferromagnetic state of a Gd$_2$CCl$_2$ monolayer, which has been fabricated in experiments. Our theoretical model together with the material predictions provide a realistic platform for the study of flat-bands and related exotic quantum phases.


\end{abstract}

\maketitle

\noindent \textbf{1. Introduction}

The properties including the electronic bands of a material are codetermined by its structure and the component elements of which the outer-shell electrons usually matter most~\cite{harrison2012electronic}. In the band theory regime, there are two limiting cases: the massless Dirac fermion in linear dispersion band~\cite{neto2009electronic} and infinitely heavy fermion in flat-band~\cite{mielke1991ferromagnetism, tasaki1992ferromagnetism}. Both of them  harbor unique and fantastic properties. The Dirac fermion has been experimentally realized in a graphene monolayer~\cite{neto2009electronic}. For the flat-band, the density of states is impressive and effects of interactions are entirely nonperturbative~\cite{mielke1991ferromagnetism, mielke1992exact, tasaki1992ferromagnetism, tasaki1998nagaoka, bodyfelt2014flatbands}. This may offer unique opportunities for the emergence of exotic quantum phases, including ferromagnetism~\cite{mielke1992exact, mielke1991ferromagnetism, tasaki1992ferromagnetism, tasaki1998nagaoka}, high-temperature fractional quantum Hall effect {~\cite{tang2011high, sun2011nearly, neupert2011fractional}},  Bose-Einstein condensation~\cite{huber2010bose}, and high-temperature superconductivity~\cite{imada2000superconductivity, peotta2015superfluidity}, Wigner crystalization~\cite{wu2007flat}.

The realization of flat-band and intriguing properties in various moir\'e superlattices has achieved great success~\cite{cao2018unconventional, cao2018correlated, chen2019evidence, ma2020topological}, where the tunability due to the two-dimensional (2D) nature plays an important role. Compared with the flat-band in a small moir\'e Brillouin zone, the intrinsic flat-band in a whole Brillouin zone of 2D atomic crystal is also attractive and may have its own merit for realization of above exotic quantum phases. To design flat-band in the 2D crystals, several theoretical models have been proposed, including single (e.g., $s$- or $d_{z^2}$-type) orbital in systems of bipartite graphs or their line graphs {~\cite{mielke1992exact, mielke1991ferromagnetism, tasaki1992ferromagnetism, tasaki1998nagaoka, bodyfelt2014flatbands, tang2011high, sun2011nearly, neupert2011fractional, huber2010bose, imada2000superconductivity, peotta2015superfluidity, lieb1989two, wang2011nearly, leykam2018artificial, chernyshev2016damped, guo2009topological, weeks2010topological, bergman2008band, yang2020gapped}}, and twofold degenerate $p_{xy}$-type orbital in honeycomb structure~\cite{wu2007flat}. Though progress has been made~\cite{slot2017experimental, hase2018possibility, aiura2017disappearance, lin2018flatbands, leykam2018artificial, drost2017topological, kang2020dirac, yin2019negative, PhysRevLettFlat}, the definitive experimental realization in 2D atomic crystal is still lacking.  In real 2D materials, the electronic structures often deviate heavily or even completely from these idealized models, when there is inevitable hybridization between the specified orbitals and other orbitals. From a more realistic consideration, the  threefold degenerate $p_{xyz}$-orbitals  exist widely and isolate well from other orbitals in ionic materials. Designing a 2D flat-band based on the degenerate $p_{xyz}$-orbitals and searching for its material realization may open a promising way.

In this paper, we propose a novel and realistic flat-band model based on the degenerate $p_{xyz}$-orbitals when they locate at the centers of octahedrons  which are closely-packed to form a 2D structure. Our theoretical analysis and first-principles calculations show that the proposed flat-bands can be realized in 1T layered materials of alkali-metal chalogenides and metal-carbon group compounds. The calculated flat-bands of these materials can be well-described by our theoretical model with physically meaningful parameters. Some of the former materials are theoretically predicted to be stable in layered structure here (e.g., K$_2$S), while some of the latter have been experimentally fabricated in previous works (e.g., Gd$_2$CCl$_2$)~\cite{schleid1994crystal, ryazanov2006la2tei2, schleid1987synthesis, lukachuk2007new}. More interestingly, the flat-band is partially filled in the heterostructure of K$_2$S and graphene layers.  The spin polarized nearly flat-band can be realized in the ferromagnetic state of a Gd$_2$CCl$_2$ monolayer. Our theoretical model together with the material predictions provide a realistic platform for the study of flat-band and related exotic quantum phases.

\noindent \textbf{2. Theoretical model}

We first discuss the tight-binding (TB) model.  In a 1T (tetragonal symmetry, octahedral coordination) layered structure, the octahedrons are closely-packed in 2D planes by sharing edges and corners, as shown in Fig. 1. We define the 2D plane formed by the centers of octahedrons as $xy$-plane. Three lattice vectors in the $xy$-plane are defined as $\vec{a}_{1}=\frac{\sqrt{3}}{2}a\hat{e}_x - \frac{1}{2}a\hat{e}_y$, $\vec{a}_{2}=a\hat{e}_y$, and $\vec{a}_{3}=-\vec{a}_2-\vec{a}_1$, where $a$ is the lattice constant and $\hat{e}_{x,y}$ are the unit vectors of the two orthogonal axes. When an anion locates at the center of an octahedron structure, its three $p$-orbitals are energetically degenerate, as shown in Fig. 1a. For convenience, we label $p_{x,y,z}$ as $p_{1,2,3}$ and align their polarization direction to the three diagonals of the regular octahedron. Cations locate at the corners of octahedrons to make the system electronically neutral. The TB Hamiltonian $H=H_0+H'$ reads
\begin{align}
H_0 &=t_0\sum_{\vec{r};i;j;k}p^\dagger_{\vec{r},i} p_{\vec{r}\pm \vec{a}_k,j}|\epsilon_{ijk}|, \\
H'&=t_1\sum_{\vec{r};i}p^\dagger_{\vec{r},i} p_{\vec{r}\pm \vec{a}_i,i} +t_2\sum_{\vec{r};i;j}p^\dagger_{\vec{r},i} p_{\vec{r}\pm \vec{a}_j,i}(1-\delta_{i,j})  \nonumber \\
&+t_3\sum_{\vec{r};i;j}(p^\dagger_{\vec{r},i} p_{\vec{r}\pm \vec{a}_j,j}+p^\dagger_{\vec{r},i} p_{\vec{r}\pm \vec{a}_i,j})(1-\delta_{i,j}) \nonumber \\
&+E\sum_{\vec{r};i;j}p^\dagger_{\vec{r},i} p_{\vec{r},j}(1-\delta_{i,j}),
\end{align}
where $i, j, k = 1-3$; $\vec{r}$ runs the locations of all anions in the $xy$-plane; $t_0$ describes the hopping between $p$-orbitals on neighboring anions when their polarization directions cross at one corner of  the octahedron, namely a cation; $t_1$ ($t_2$) describes the neighboring hopping between the same $p$-orbitals when they are (not) perpendicular to the connection direction $\vec{a}_i$; $t_3$ is the hopping between different $p$-orbitals when only one of them is perpendicular to the connection direction $\vec{a}_i$. The values can be estimated as
$t\approx -\sum\int p^*_i(r)v_\text{nc}p_j(r-\vec{a}_k){\rm d}r$, where $v_\text{nc}$ are the atomic potentials of the nearest cations. The schematic diagrams of hoppings along $\vec{a}_1$ direction are shown in Fig. 1c. The gray solid circles represent the $v_\text{nc}$ that are most contributive. It is obvious that $t_0$ is the only one case that wave functions $p_i$ and $p_j$, and atomic potentials $v_\text{nc}$ overlap directly. Thus, the amplitude of $t_0$ is expected to be apparently larger than $t_1$, $t_2$, and $t_3$. $E$ describes the crystal field when the octahedrons are arranged in the 2D plane and the effect of structure adjustment, which is tunable via adjusting the structure or varying the chemical environment.
\begin{figure}
\includegraphics[width=0.9\columnwidth]{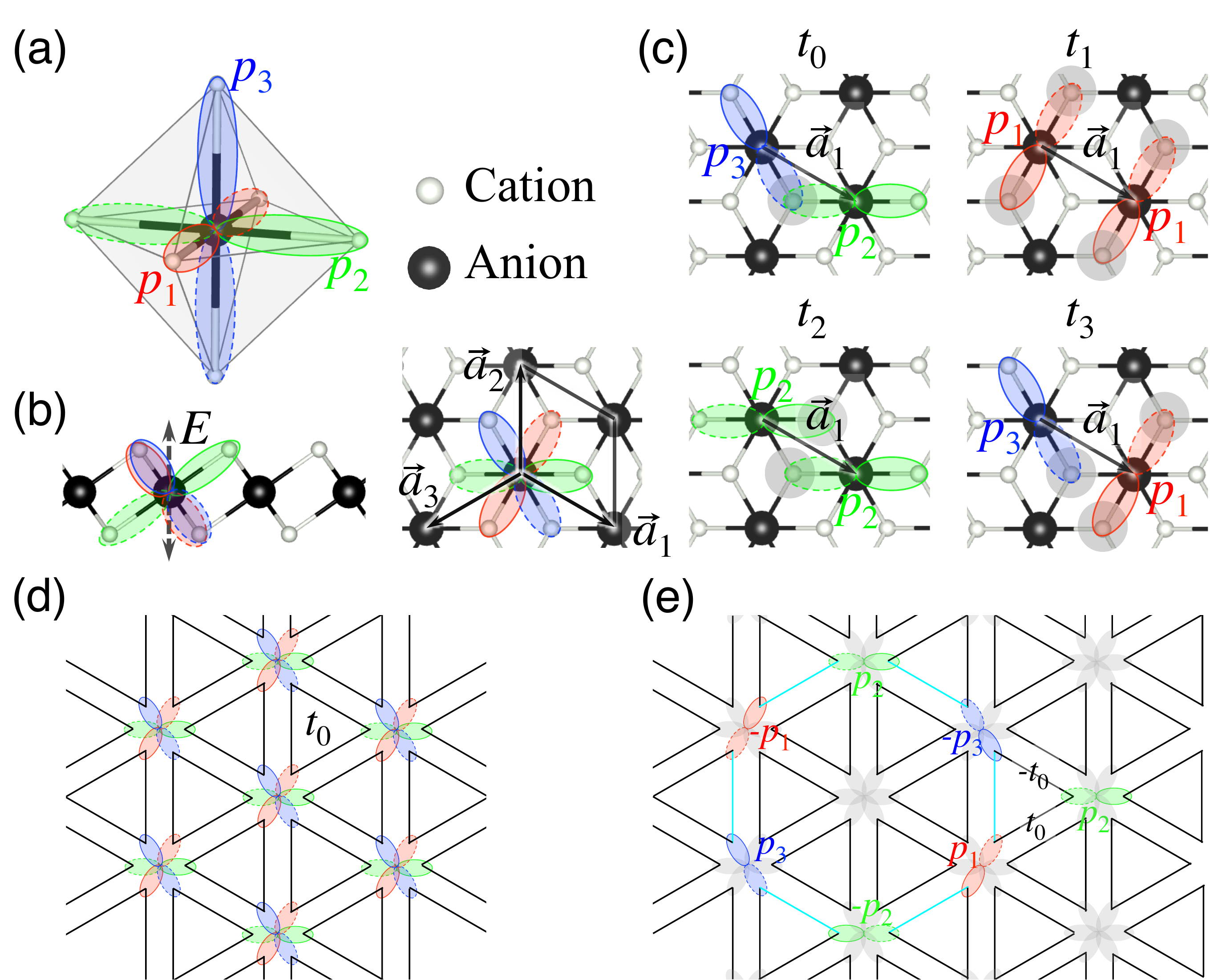}
\colorcaption{The structure and hoppings between $p$-orbitals in a 1T monolayer. (a) An octahedron formed by six cations with an anion at the center. The $p_{1,2,3}$-orbitals on the anion are represented by red, green, and blue dumbbells, respectively. For a $p$-orbital, the dashed and solid edges of ellipses represent the opposite phases of its wave function. (b) Schematic diagram of the effective crystal field $E$. The black rhombus represents one unit cell and black arrows are the lattice vectors $\vec{a}_{1,2,3}$. (c) Schematic diagrams of the hoppings between $p$-orbitals along $\vec{a}_1$ direction. The grey solid circles represent the atomic potentials $v_\text{nc}$ of nearest cations that contribute to the hopping term most. (d) Schematic diagram of the lattice model of $H_0(t_0)$. (e) A compact localized state composed of 6 connected orbitals highlighted in color. The loop formed by these orbitals is colored in cyan.}
\label{fig1}
\end{figure}

It is of practically guiding significance to consider a limiting case when only $t_0$ has nonzero value with $t_1=t_2=t_3=E=0$ and $H=H_0$. Fig. 1d shows a schematic diagram of the lattice structure of $H_0(t_0)$. In momentum space, we define the three component basis as
\begin{equation}
\psi(\vec{k})=[p_1(\vec{k}),p_2(\vec{k}),p_3(\vec{k})]^{\rm T}.
\end{equation}
Then $H(\vec{k})$ takes the matrix form as
\begin{equation}
2t_0\left(
\begin{array}{ccc}
0 & \text{cos}(k_3) & \text{cos}(k_2) \\
\text{cos}(k_3) & 0 & \text{cos}(k_1) \\
\text{cos}(k_2) & \text{cos}(k_1) & 0  \\
\end{array}
\right),
\end{equation}
where $k_i=\vec{k}\cdot\vec{a}_i$ is defined in the 2D Brillouin zone. The band structure contains three bands $E_{1,2}=t_0\pm t_0\sqrt{3+2\sum_{i}{\text{cos}(2k_i)}}$ and $E_3=-2t_0$, as shown in Fig. 2a. Interestingly, the $E_3$  band is totally flat over the entire 2D Brillouin zone . On the other hand, the $E_{1,2}$ bands are dispersive exhibiting eight Dirac cones at K and $\frac{1}{2}$K points in the first Brillouin zone . The bandwidth of $E_{1,2}$ is determined by the amplitude of $t_0$. The dispersive band $E_2$ and flat-band $E_3$ are energetically degenerate at $\Gamma$ and M points. The totally flat-band is attributed to the frustrated hopping nature~\cite{bergman2008band}. The loop in Fig. 1e represents a compact localized state and the frustrated hopping to a neighbor orbital.

With regard to spin-orbit coupling (SOC), we consider the original atomic form: $H_\text{SOC}=\lambda\vec{L}\cdot\vec{\sigma}$~\cite{liu2011low, liu2013flat}. It reads
\begin{equation}
H_\text{SOC}=\lambda\left(
\begin{array}{ccc}
 & -i\sigma_3 &  i\sigma_2 \\
 i\sigma_3 & 0 &  -i\sigma_1 \\
 -i\sigma_2 &  i\sigma_1 & 0  \\
\end{array}
\right),
\end{equation}
where $\sigma_{1,2,3}$ are Pauli matrices~\cite{liu2011low, liu2013flat}. As shown in Fig. 2b, the band degeneracy lifts at the high symmetry $\Gamma$, M, and K points due to the effects from SOC. A global gap occurs, which isolates the flat-bands from other bands. The flat-band becomes slightly dispersive after taking finite $H_\text{SOC}$ and $H'$ into consideration.

\begin{figure}[tb]
\includegraphics[width=1\columnwidth]{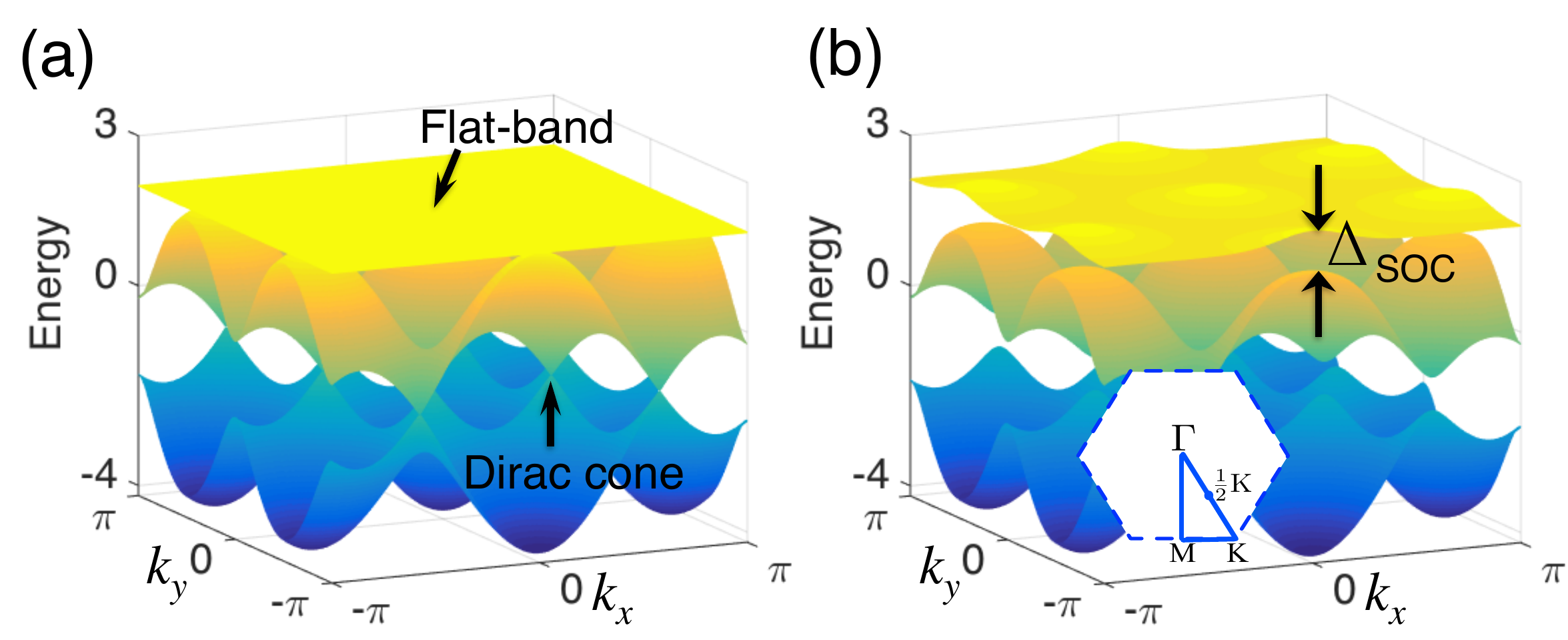}
\colorcaption{Band dispersion of the three degenerate $p$-orbitals with the parameters: (a) $t_0=-1$, $t_1=t_2=t_3=E=\lambda=0$; (b) $t_0=-1$, $t_1=t_2=t_3=E=0$, $\lambda=0.2$. In (a), the band $E_3$ is completely flat, while $E_{1,2}$ exhibit Dirac cones at K points and $\frac{1}{2}$K points. In (b), the band degeneracy lifts at the high symmetry $\Gamma$, M, and K points due to the effects from SOC when $\lambda\neq0$. $\Delta_\text{SOC}$ in (b) is the direct gap between the nearly flat-band $E_3$ and band $E_2$ at $\Gamma$ and M points. The inset in (b) is the Brillouin zone  and high-symmetry path.}
\label{fig2}
\end{figure}

\noindent \textbf{3. Material realization}

With the help of first-principles calculations, we predict that two classes of layered materials may harbor nearly flat-bands: 1T layered alkali-metal chalogenides and metal-carbon group compounds. To be specific, we give 1T  dipotassium monosulfide (K$_2$S) monolayer and digadolinium monocarbide dichloride (Gd$_2$CCl$_2$) monolayer as two examples.

\begin{table}[tb]
\caption{Energy comparison of alkali-metal chalogenides in different structures. The first row (column) lists the symbol of chalcogens (alkali metals). $E_\text{1T}-E_\text{layered}$ is the exfoliation energy of a 1T monolayer from its layered bulk.  The energy difference $E_\text{2H}-E_\text{1T}$ is the relative structural stability of a 1T monolayer to its 2H counterpart.}
\centering
\begin{tabular}{p{0.12\linewidth}<{\centering}p{0.29\linewidth}<{\centering}p{0.12\linewidth}<{\centering}p{0.12\linewidth}<{\centering}p{0.12\linewidth}<{\centering}p{0.12\linewidth}<{\centering}}
\hline
 &
Energy difference (eV/unit cell)                            &      O    &     S     &     Se    &   Te       \\
\hline
\multirow{2}*{K} &  $E_\text{1T}-E_\text{layered}$  &   0.01    &   0.18   &   0.22    &  0.27  \\
       			&  $E_\text{2H}-E_\text{1T}$        &   0.51    &   0.45   &   0.42   &   0.37  \\
\hline
\multirow{2}*{Rb}&  $E_\text{1T}-E_\text{layered}$  &   0.09    &   0.30   &   0.34    &  0.42  \\
       			&  $E_\text{2H}-E_\text{1T}$        &   0.43    &   0.39   &   0.37   &   0.33  \\
\hline
\label{tb:Reference Vectors1}
\end{tabular}
\end{table}

Our density functional theory (DFT) calculations are carried out using the Vienna \textit{ab initio} simulation package ({\footnotesize{VASP}})~\cite{VASP1996}, where the projector augmented plane wave (PAW) method~\cite{blochl1994projector, kresse1999ultrasoft} is adopted, and the generalized gradient approximation (GGA) in the framework of Perdew-Burke-Ernzerhof (PBE)~\cite{perdew1996generalized} is chosen for the exchange-correlation interaction. The magnetic and electronic properties of the Gd$_2$CCl$_2$ system are calculated based on the range-separated Heyd-Scuseria-Ernzerhof 2006 (HSE06) hybrid functional~\cite{heyd2003hybrid, krukau2006influence}. A specific semi-empirical scheme (DFT-D2)~\cite{grimme2006semiempirical} is used to treat the van der Waals (vdW) type interaction. Phonon spectrum are obtained using the  {\footnotesize{PHONOPY}} code~\cite{Phonopy2015}.

Table~I lists the comparisons of total energies of alkali-metal chalogenides in different structures.  All the 1T monolayers are energetically more favored than their 2H  (hexagonal symmetry, trigonal prismatic coordination) counterparts, which benefits the synthesis of high quality 2D samples. The 2D monolayer would be epitaxially grown and further stabilized on proper substrates~\cite{hong2017atomic, zhu2017multivalency, ding2020exploring, ding2019signature, zeng2017half}.  The structural stabilities of these 1T materials are also checked by their nonnegative phonon spectra.

\begin{figure}[tb]
\includegraphics[width=0.9\columnwidth]{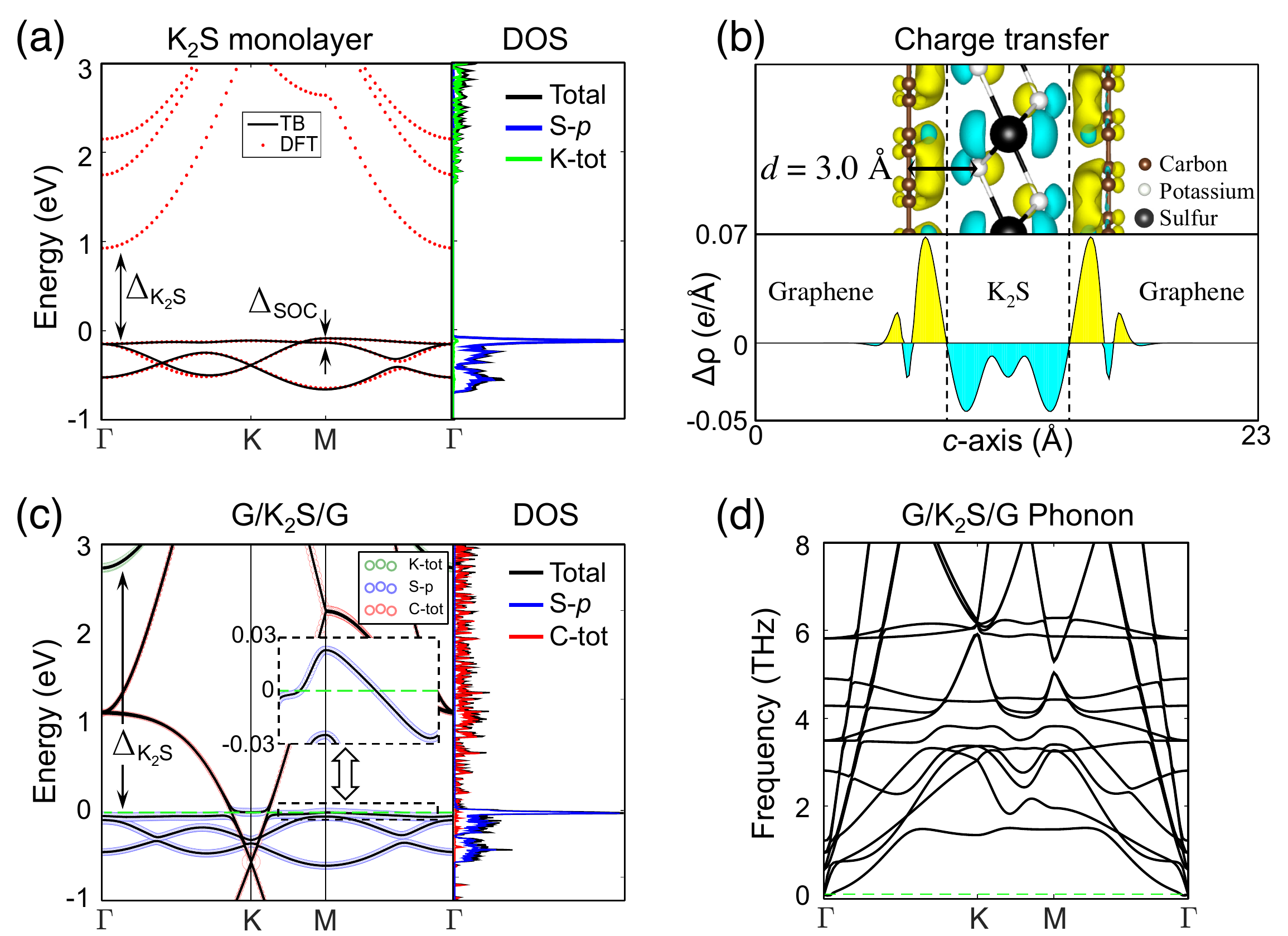}
\colorcaption{Electronic properties and phonon spectra of 1T K$_2$S systems. (a) Electronic bands of a freestanding 1T K$_2$S monolayer. Filled dots and solid curves represent the data from DFT calculations and TB fittings, respectively. The fitting parameters are: $t_0=-83$, $t_1=-19$, $t_2=13$, $t_3=5$, $E=20$, and $\lambda=20$ meV. The highest valence bands are nearly flat with bandwidth of 55 meV. $\Delta_\text{SOC}=54$ meV is the SOC gap at M point. (b) Charge distribution $\Delta \rho$ in the G/K$_2$S/G heterostructure and the projection along $c$-axis. The isosurface level is 0.003 $e$/\AA$^3$. Positive and negative values are colored in yellow and cyan, respectively.  (c) Orbital projected electronic bands of the G/K$_2$S/G heterostructure. The contributions from the K atoms (K-tot), C atoms (C-tot), and $p$-orbitals of S atoms (S-$p$) are denoted by hollow circles in green, red, and blue color, respectively. The inset is a zoom-in view of the partially filled nearly flat-band. The right panels in (a) and (c) are the DOS. $\Delta_{\text{K}_2\text{S}}$ is the band gap of the K$_2$S monolayer.  (d) Phonon spectra of the G/K$_2$S/G heterostructure.}
\label{fig3}
\end{figure}

Some of the above 1T monolayered alkali-metal chalogenides harbor nearly flat-bands. As an example, the electronic band structures of the 1T K$_2$S monolayer are shown in Fig. 3a. Our TB model fits well with the DFT calculated electronic bands with the parameters shown in the caption. The highest valence band is nearly flat with bandwidth of 55 meV. The atomic structure refers to Fig. 1. The optimized lattice constant is: $a=5.01 ~\text{\AA}$.

Due to the 2D nature, monolayered materials can be conveniently stacked to form heterostructures. The charge transfer naturally takes place between the 2D materials when their work functions are different. Here, to explicitly show the doping effects on the flat-band materials, we theoretically study a G/K$_2$S/G heterostructure in which a K$_2$S layer is sandwiched in between two graphene layers. The lattice mismatch between a K$_2$S unit cell and  2$\times$2 graphene supercell is less than 2\%. In Fig. 3b, the charge distribution is defined as $\Delta\rho=\rho_{\text{G/K}_2\text{S/G}}-\rho_{\text{K}_2\text{S}}-\rho_\text{G}$, where $\rho_{\text{G/K}_2\text{S/G}}$, $\rho_{\text{K}_2\text{S}}$, and $\rho_\text{G}$ are the charge densities of the G/K$_2$S/G heterostructure, K$_2$S monolayer, and graphene layers, respectively. The calculated electrons transfer from the K$_2$S layer to graphene layers is 0.13 electrons per unit cell. This evident value is mainly attributed to the large difference $\sim$3.0 eV in their work functions. Fig. 3c shows the orbital projected electronic bands. The nearly flat-band is partially filled and the Dirac points of graphene is shifted down below Fermi level due to the charge transfer effect. The filling level can be further tuned via electric gating due to the 2D nature. On the other hand, the nearly flat-band is almost intact from hybridization with other bands. The nonnegative phonon spectra in Fig. 3d verifies the structural stability. In addition, it notes that the band gap $\Delta_{\text{K}_2\text{S}}$  of a K$_2$S monolayer is environmentally sensitive largely due to its ionic nature, as comparatively shown in Fig. 3a and c.

As another large family of materials, compounds of metal and carbon group elements can also adopt a 1T layered structure,  some of which have been fabricated in experiments~\cite{schleid1994crystal, ryazanov2006la2tei2, schleid1987synthesis}. We present a 1T Gd$_2$CCl$_2$ monolayer as a typical example. Fig. 4a shows the structure of a 1T Gd$_2$CCl$_2$ monolayer. In the 1T phase, one C atom sites at the inversion symmetry point of an octahedron formed by six Gd atoms. Different magnetic orders in a $\sqrt{3}\times \sqrt{3}$ supercell are compared to find the ground state. Based on our HSE06 calculations, the Gd atoms prefer a ferromagnetic order. The ferromagnetic order could be further stabilized via applying external magnetic field~\cite{deng2020quantum}. Fig. 4b and c show the spin polarized band structures. Similar to the case of K$_2$S monolayer, the highest valence band of the spin up electron in Gd$_2$CCl$_2$ is very flat with a bandwidth of 195 meV. From the orbital projected  orbital projected density of states (DOS), it is clear that the three highest valence bands are mainly contributed from the $p$-orbitals of C anions.

\begin{figure}[tb]
\includegraphics[width=1\columnwidth]{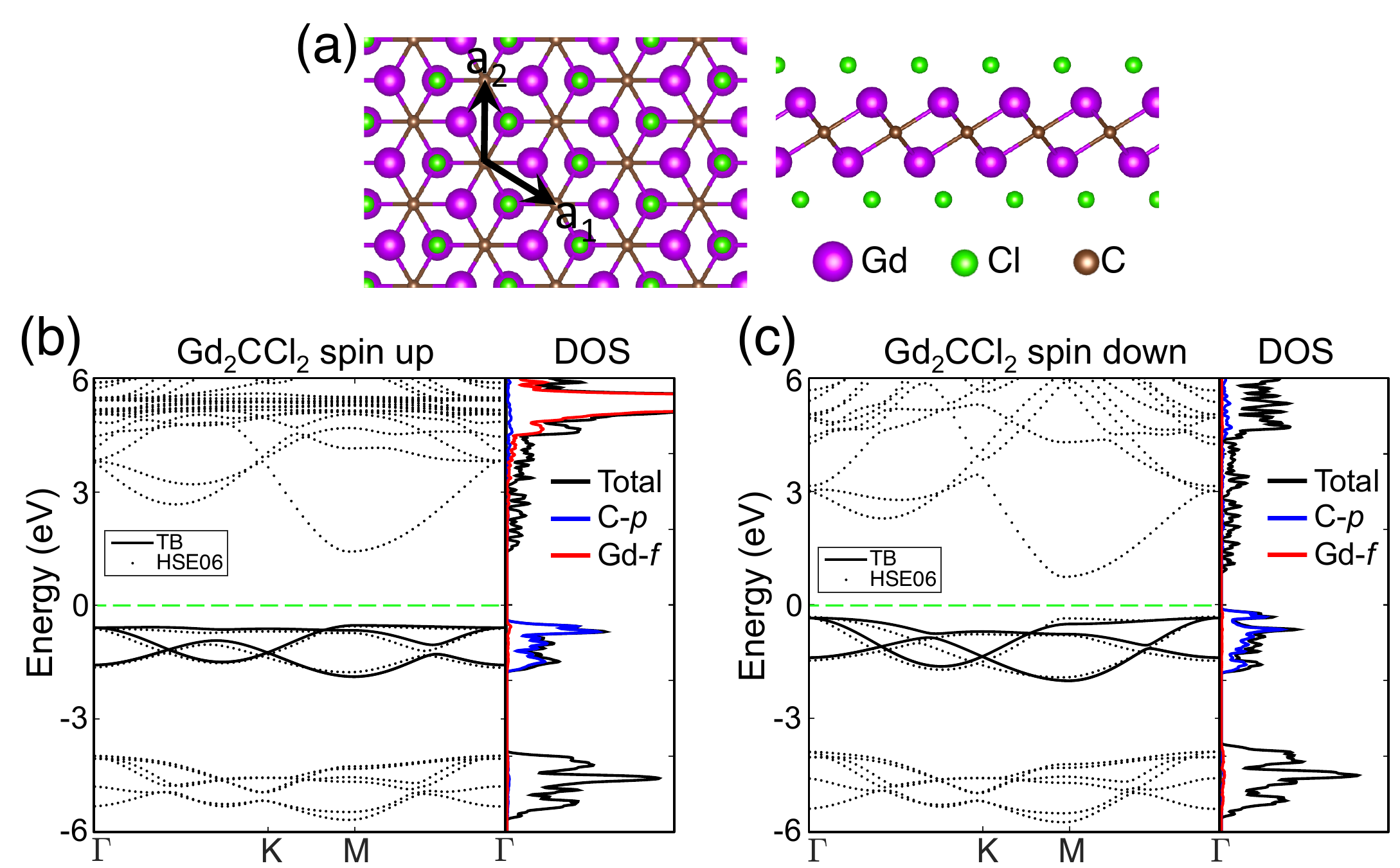}
\colorcaption{The structure and spin-resolved electronic bands of a freestanding 1T Gd$_2$CCl$_2$ monolayer. (a) The top view and side view of the structure. The electronic bands along high-symmetry paths are shown in (b) for spin up and (c) for spin down, respectively.  Filled dots and solid curves represent the data from HSE06 calculations and TB fittings, respectively. The right panels in (b) and (c) are the orbital projected DOS. The black, red, and blue curves are the total DOS, contributions from the $p$-orbitals of C atoms (C-$p$) and $f$-orbitals of Gd atoms (Gd-$f$), respectively.}
\label{fig4}
\end{figure}

 \noindent \textbf{4. Discussion and conclusion}

In summary, a novel TB model with (nearly) flat-bands was proposed based on the  threefold degenerate $p$-orbitals in 2D materials. As concrete examples, our calculations showed that a Gd$_2$CCl$_2$ monolayer harbors a spin polarized nearly flat-band in its ferromagnetic phase, and the flat-band is partially filled in the G/K$_2$S/G heterostructure. Our theoretical model together with the material predictions provide a realistic platform for the study of flat-band and related exotic quantum phases.

 \noindent \textbf{Conflict of interest}

The authors declare that they have no conflict of interest.

\noindent \textbf{Acknowledgments}

The authors thank Wei Qin and Zhenyu Zhang for helpful discussions. This work was supported by the National Basic Research Program of China (2015CB921102 and 2019YFA0308403), the National Natural Science Foundation of China (11674028 and 11822407), the Strategic Priority Research Program of Chinese Academy of Sciences (Grant No. XDB28000000), and China Postdoctoral Science Foundation (2020M670011).

 \noindent \textbf{Author contributions}

Jiang Zeng and X. C. Xie conceived the idea and supervised the project. Jiang Zeng did the theoretical analysis and  first-principles calculations.  Jiang Zeng, Ming Lu, Hua Jiang, Haiwen Liu, and X. C. Xie analyzed the data and wrote the manuscript. All authors contributed to scientific discussion of the manuscript.

\end{document}